\documentclass{article}

\usepackage{microtype}
\usepackage{graphicx}
\usepackage{subfigure}
\usepackage{booktabs} % for professional tables

\usepackage{hyperref}

\usepackage[accepted]{icml2024}

\usepackage{amsmath}
\usepackage{amssymb}
\usepackage{mathtools}
\usepackage{amsthm}

\usepackage[capitalize,noabbrev]{cleveref}

\theoremstyle{plain}

\theoremstyle{definition}

\theoremstyle{remark}

\usepackage[textsize=tiny]{todonotes}

\icmltitlerunning{NaNa and MiGu: Semantic Data Augmentation Techniques to Enhance Protein Classification in Graph Neural Networks}

\begin{document}

\twocolumn[
\icmltitle{NaNa and MiGu: Semantic Data Augmentation Techniques to Enhance \\ Protein Classification in Graph Neural Networks}

\icmlsetsymbol{equal}{*}

\begin{icmlauthorlist}
\icmlauthor{Yi-Shan Lan}{nthu}
\icmlauthor{Pin-Yu Chen}{ibm}
\icmlauthor{Tsung-Yi Ho}{cuhk}
\end{icmlauthorlist}

\icmlaffiliation{nthu}{Department of Computer Science, National Tsing Hua University, Hsinchu, Taiwan}
\icmlaffiliation{cuhk}{Department of Computer Science and Engineer, The Chinese University of Hong Kong, Shatin, Hong Kong}
\icmlaffiliation{ibm}{IBM Research, New York, USA}

\icmlcorrespondingauthor{Tsung-Yi Ho}{
tyho@cse.cuhk.edu.hk}

\icmlkeywords{Machine Learning, ICML}

\vskip 0.1in
]

\newcommand{\YSNOTE}[1]{{\color{blue}[YSNOTE: #1]}}
\newcommand{\SYNOTE}[1]{{\color{green}[SYNOTE: #1]}}
\newcommand{\red}[1]{{\color{red}#1}}

\newcommand{\PY}[1]{{\color{blue}PY: #1}}
\newcommand{\PYB}[1]{{\color{blue}[PY: #1]}}

\printAffiliationsAndNotice{\icmlEqualContribution} % otherwise use the standard text.

\begin{abstract}
Protein classification tasks are essential in drug discovery. Real-world protein structures are dynamic, which will determine the properties of proteins. However, the existing machine learning methods, like ProNet \cite{wang2022learning}, only access limited conformational characteristics and protein side-chain features, leading to impractical protein structure and inaccuracy of protein classes in their predictions. In this paper, we propose novel semantic data augmentation methods, Novel Augmentation of New Node Attributes (NaNa) and Molecular Interactions and Geometric Upgrading (MiGu) to incorporate backbone chemical and side-chain biophysical information into protein classification tasks and a co-embedding residual learning framework. Specifically, we leverage molecular biophysical, secondary structure, chemical bonds, and ionic features of proteins to facilitate protein classification tasks. Furthermore, our semantic augmentation methods and the co-embedding residual learning framework can improve the performance of GIN \cite{GIN} on EC and Fold datasets \cite{EC, SCOP} by 16.41\% and 11.33\% respectively. Our code is available at \url{https://github.com/r08b46009/Code_for_MIGU_NANA/tree/main}.
% For example, our method outperform the state-of-the-art protein graph model, ProNet \cite{wang2022learning}, by 4.32\% and 2.78\% on EC and Fold datasets
%Consequently, we contribute a biophysics and chemical-based data augmentation method to enhance protein representation learning for the most popular network architectures.
%Shed light on embedding accurate dynamic structures and chemical information of proteins into representation learning for future research.

\begin{figure*}[ht]
\begin{center}
\centerline{\includegraphics[width=0.98\linewidth]{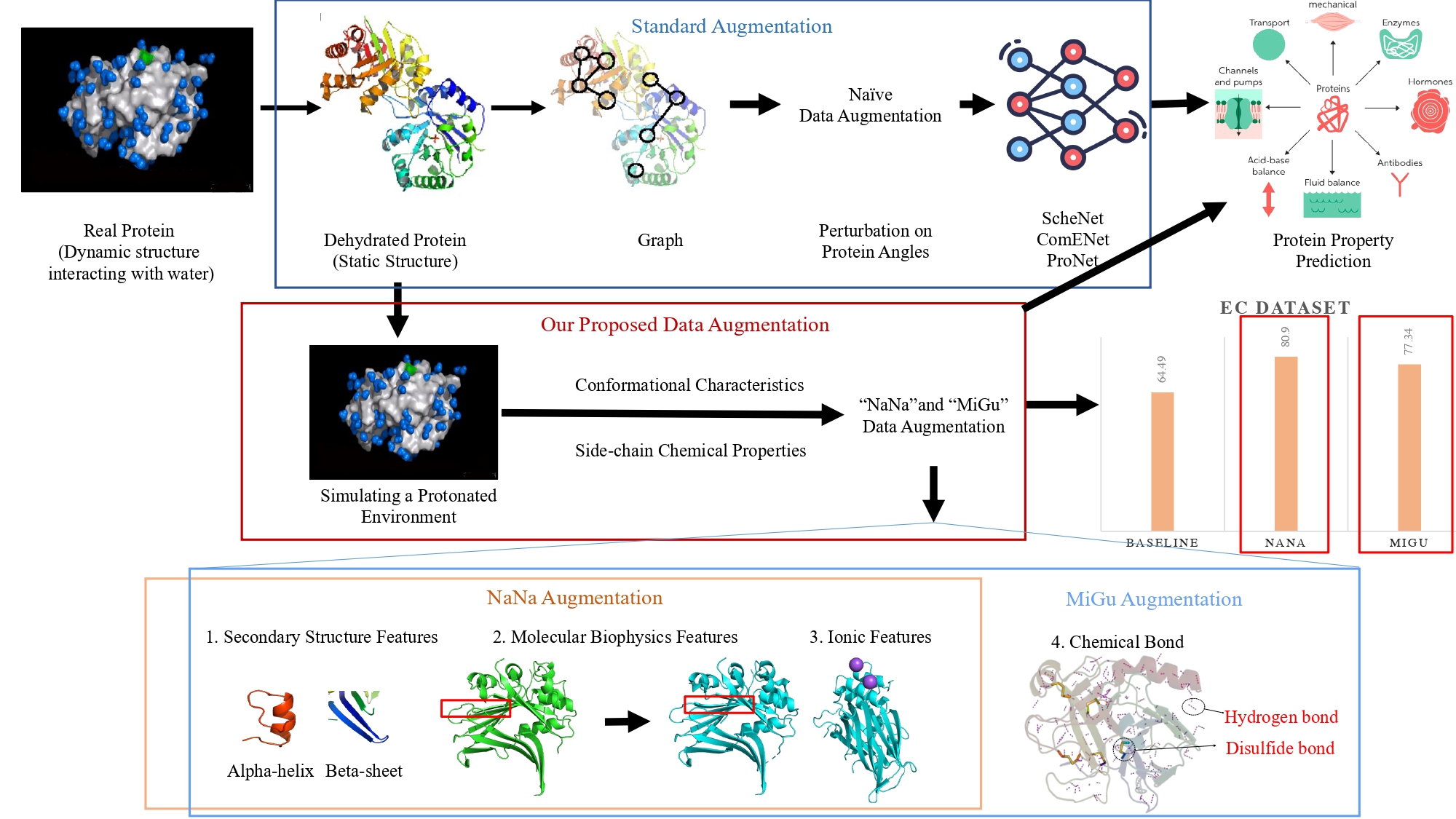}}
    \caption{Schematic representation of the conventional graph neural network training process compared to our semantic data augmentation pipeline (NaNa and MiGu) for protein classification tasks. Our method introduces a protonation step to model static structures dynamically, inducing conformational changes that facilitate the extraction of side-chain biophysical properties. We also incorporate secondary structural features obtained through the DSSP algorithm \cite{kabsch1983dictionary} to enrich our dataset. The integration of ionic types further extends the dataset to encapsulate a more complete representation of structural information \cite{ion}, aiming to improve the predictive performance as demonstrated in the EC dataset \cite{EC}. 
    % \PYB{1. Change Our Proposal to "Our Proposed Data Augmentation", Solved 2. I recall some features correspond to NaNa and Some correspond to MiGU. Could you categorize them with the features you listed?, Solved 3. Remove the circle of Conformal ..., Solved 4. Change Our Data Augmentation to "NaNa and MiGu", Solved 5. It's better to say what features belong to the category of NaNa and what belongs to MiGu, Solved} 
    In considering the different protein classification tasks, we design two different semantic protein augmentation, NaNa and MiGu. NaNa incorporates important secondary structure, molecular biophysics, and ionic features to achieve semantic augmentation with biochemical and biophysical properties in proteins, leading to remarkable performance in Fold dataset. On the other hand, to achieve semantic augmentation with molecular interaction information, MiGu augmentation extends the NaNa augmentation approach by including bond-type features, offering a more comprehensive augmentation for protein classification tasks, like Superfamily dataset.    
    % Also, can they be used together? Why or why not? 
    }

\label{fig:icml_historical}
\end{center}
\vskip -0.2in
\end{figure*}

\end{abstract}

\section{Introduction}
\label{submission}

Protein classification is a pivotal task in drug discovery, including classification by Enzyme Commission (EC) numbers and the Structural Classification of Proteins (SCOP) database \cite{EC, SCOP}. 
% \PYB{Need to give full names for EC and SCOP, Solved}
For example, researchers discovered that predicting the EC number of a new mitochondrial decarboxylase would offer a better understanding of the underlying mechanisms of Parkinson’s disease. This helps develop new therapies for neurotransmitter regulation \cite{mito}.
Additionally, predicting SCOPe classes can provide insights into the structural evolution of proteins. For example, researchers studied the structure of the HIV capsid protein and classified it into the SCOP category. This classification revealed that the capsid protein belonged to a specific fold unique to certain retroviruses. \cite{HIV} Therefore, it is crucial to understand the relationship between protein structures and their classes \cite{erlanson2016twenty}.

Recently, previous work has focused on more advanced graph neural network (GNN) models to achieve remarkable progress in protein classification tasks from graph representations of proteins, including ProNet \cite{wang2022learning}, ComENet \cite{wang2022comenet}, and SchNet \cite{schutt2018schnet}. However, most of them ignore the measurement error between the static protein structure and the real-world structure caused by the dehydration and low temperature of the protonated process, which is the preprocessing process of protein structure measurement. They only leverage the static protein structure graph and naive data augmentation method to achieve state-of-the-art prediction performance. For instance, ProNet \cite{wang2022learning}, ComENet \cite{wang2022comenet}, and SchNet \cite{schutt2018schnet} incorporate rotational features as geometric information to capture the different conformation in structures. However, it can only capture limited conformational characteristics, causing unreasonable protein structure because of the missing force field and side-chain biophysical prior knowledge. Besides, existing representation learning methods for protein structures were limited to only amino acid types, discharging essential ionic information for understanding protein structures \cite{ion}.

To provide more realistic features and diverse context for enriching training samples and to mitigate the distribution shift between the augmented and the real datasets, the machine learning community has made considerable efforts in semantic data augmentation to synthesize realistic backgrounds or context to diversify training images \cite{seman}. For instance, DA-Fusion \cite{da_fusion} leverages diffusion models to synthesize meaningful image backgrounds while keeping the subject unchanged. In addition,  \cite{auto_semantic_lang} develops a semantic consistent data augmentation for language models and code summarization tasks. However, the investigation of protein representation learning and classification tasks remains absent.

Inspired by semantic visual and language augmentation in different domains, we propose a novel semantic data augmentation approach for protein structures, called NaNa and MiGu, to synthesize more realistic protein information for various classification datasets. Our semantic-based protein structure augmentation consists of a biophysics data augmentation 
% \PYB{why residual learning is considered in the augmentation framework? Does the augmentation not work without this residual learning? Better separate these two contributions; okay, I will separate the content into the next paragraphs; thanks!} 
for the augmented features. Moreover, our semantic protein augmentation provides semantic context
% \PYB{you mean paralleled compute? I typed wrong. I did not utilize parallel computing, but I mentioned in 3.2.3 that calculation would be faster with parallel techniques.} 
and requires only about 4 seconds of computational time for each sample with only 3\% computational resource of Intel i7-9700K CPU and 614 MB memory. 
% \PYB{using what GPU?, Sorry again, I typed wrong. We utilized only CPU to generate augmented features} 
Specifically, we use protein dynamic simulation techniques to generate semantic features, like AMBER \cite{case2005amber} and PropKa \cite{sondergaard2011improved}, to simulate the protonated states of protein to extract accurate biophysical features, leading to more realistic protein structure support for protein classification tasks.

Additionally, we have engineered a high-efficiency residual network framework tailored for training Graph Neural Networks (GNNs). This residual learning structure delivers the additional features into deeper layers of deep models, causing better prediction accuracy in protein classification tasks and quicker convergence in network training. In the experiments, we implement these residual connections on the layers of deep Message Passing Neural Networks (MPNNs) \cite{MPNN}, Graph Convolutional Networks (GCNs) \cite{RN65}, and Graph Isomorphism Networks (GINs) \cite{GIN}. Our design specifically caters to the seamless incorporation of semantic augmentation features, including node attributes that emphasize biophysical and chemical properties within protein structures, showing remarkable performance improvements in classification tasks.

On a high level, our data augmentation methods consist of two main attributes, including new node and edge attributes. The node attributes contain four biophysical sub-features: node coordinates, molecular biophysics features, secondary structure properties, and node type features. On the other hand, the edge attributes are the predicted potential chemical bonds between atoms. On the other hand, we also propose a co-embedding residual learning architecture to inject co-embedding into deeper layers to achieve better performance than naive residual architecture.

Following our proposed semantic data augmentation illustrated in \cref{fig:icml_historical}, the model could fix missing biophysical information among the protein, improving the accuracy of prediction in baselines based on static structural datasets. Our experiments show that in the protein functionality prediction tasks, like enzyme commission number classification dataset (EC dataset) \cite{EC}, we outperformed ProNet and ComENet \cite{wang2022learning, wang2022comenet} by 4.32\% and 11.62\%. Additionally, in  FOLD datasets \cite{SCOP}, we also outperform ProNet and ComENet \cite{wang2022learning, wang2022comenet} by 2.78\%  and 13.62\%, respectively. These improvements showed that our features could significantly improve the protein functional and evolutionary classification tasks of existing GNN models.

In summary, we conclude our main contributions as follows:

% \begin{itemize}
% \item \textbf{Integrating Biochemistry and Dynamic and Geometric features on baseline models, surpassing state-of-the-art methods}: To our best knowledge, we provide a pilot study on incorporating side-chain biophysical
% and backbone interactions for protein functional predictions, surpassing state-of-the-art methods. 

\begin{itemize}
\item \textbf{Proposing semantic protein structure data augmentation techniques based on biophysic information}: Our work pioneers the semantic protein data augmentation with biophysic prior knowledge, including dynamic and geometric aspects of protein structures, into existing GNN baseline models. In addition, we propose two semantic protein structure data augmentation methods, NaNa and MiGu. These semantic augmentations significantly advance the predictive accuracy of these models, surpassing current state-of-the-art methods of protein classification with only a one-time 4-second computational time for each data sample with an Intel i7-9700K CPU. 
% \PYB{with what compute resource, Solved}. 

\item \textbf{Exploring the influence of biochemistry and dynamic features with leave-one-out analysis}: In our research, we conduct systematic research on feature analysis to unravel the significance of specific chemical and biophysical properties derived from protonated structures. This analysis uniquely incorporates a spectrum of factors, including molecular biophysics features, secondary structure features, chemical bonds, and ionic types, thereby providing feature importance analysis into the computational prediction of protein functionalities. For example, our leave-one-out analysis revealed that certain secondary structure features play a more critical role than previously understood in EC and SCOPe datasets. This insight challenges conventional thinking in the field and opens new avenues for exploring protein classification tasks.
% \item \textbf{Creating residual network learning architectures to achieve shorter convergence time with node attributes}: 
% We created a new residual network learning architecture \cite{zhou2023co} based on the MPNN and GIN model, to better extract the essential biophysical and side-chain chemical features for protein function prediction in most datasets.  

\item 
\textbf{Developing efficient residual network architectures for accelerated GNN training:}
We introduce a novel residual network learning architecture for delivering messages into deeper layers, leading to significant accuracy improvement for protein classification tasks and faster convergence speed. In the experiment, we also verify this architecture on popular graph neural networks, like MPNN \cite{MPNN}, GCN \cite{RN65}, and GIN \cite{GIN}. This architecture is specifically tailored to efficiently process node attributes, focusing on biophysical and chemical features in protein structures, and has shown remarkable performance improvements in various datasets.

\end{itemize}

\section{Related Work}

\subsection{Protein Structure Representation}

Recent research has delved into representation learning for small molecules possessing 3D structures \cite{RN61, Protein, wang2022learning, schutt2018schnet, wang2022comenet}. GraphQA \cite{RN75}  introduces a representation learning approach, including node features representing various amino acid characteristics, dihedral angles, surface accessibility, and secondary structure types in the context of learning tasks. It recognizes the importance of capturing both the primary structure (residue sequence), secondary structure, and tertiary structure (spatial arrangement) of proteins for effective representation learning \cite{RN75}. In contrast, IEConv addresses the multi-level structure of proteins and the need to capture various structural invariances. It recognizes that proteins comprise primary, secondary, tertiary, and quaternary structures, each contributing to their functions \cite{RN61}. 
Representing proteins with 3D structures is challenging, especially when dealing with structurally unstable regions like dynamic nature \cite{RN63}. Traditional methods cannot capture their dynamic nature. To address these challenges, introducing chemical information, protonation states, hydrogen bonding patterns, and surface accessibility offers a potential solution. This chemical data provides valuable insights into how proteins behave and interact, bridging the gap between their structural flexibility and functional diversity \cite{hbond, surface}.

\subsection{Harnessing Graph Neural Networks for Protein Structure Classification}

GNNs have gained significant importance in protein structure classification \cite{GNN_P}. Protein structures naturally exhibit graph-like characteristics, with interactions between amino acid residues represented as edges in a graph. This inherent graph structure is unsuited for traditional linear models, making using GNNs a logical choice \cite{protein_structure}. In protein structure classification, various types of GNNs are employed. For example, GCNs are widely used GNN models that perform convolution operations on neighboring nodes, making them suitable for capturing local features in protein structures \cite{RN65}. GraphSAGE is a GNN model that samples neighbor node features, making it efficient for handling large-scale graph data often encountered in protein structure classification tasks \cite{GRAPHSAGE}. GIN are GNN models based on graph isomorphism, enabling them to comprehensively capture both local and global characteristics of protein structures, thereby excelling in classification tasks \cite{GIN}. MPNNs adapt to various graph structures and tasks, offering versatility for diverse applications, including protein structural learning. \cite{wang2022learning} They iteratively update node information, capturing local and global graph features.

To illustrate the use of GNNs in protein structure classification, consider representing different amino acid residues as nodes in a graph and their interactions as edges. Additionally, GCNs can be employed to capture local features \cite{RN65}, GraphSAGE for efficient processing of large datasets \cite{GRAPHSAGE}, GINs for a holistic consideration of global and regional characteristics, enhancing the accuracy of protein structure classification \cite{GIN}, or MPNNs for capturing both local and global features by analyzing protein structures and functions \cite{wang2022learning}. 

\subsection{Relationship between Chemical information and Protein Classification}
\label{author info}

Although geometry-based models have been predominant in this field, our research aims to broaden the scope by incorporating chemical insights into protein structure classification. Previous studies have revealed the significant impact of protonation on protein structure \cite{RN77, RN66}; by adjusting the electrostatic environment with a positive charge, we could influence protein structure and function to get different states of protein conformation, showing more diversity of functionality \cite{elec}. Moreover, protonation could help us capture the potential pivotal hydrogen bonds, often associated with various chemical reactions and instrumental in comprehending protein function \cite{hbond, RN79}. In addition, the hydrogen bonds could form a secondary structure, facilitating the protein function and motivating many enzymatic activities \cite{RN100}. To stabilize the secondary structure, metal ions are essential to stabilize electrostatic interactions \cite{RN47, RN69}. The information above is pivotal when simulating the protein structure to induce their potential function.
By integrating critical chemical properties such as protonation data \cite{RN66}, determination of chemical bond types \cite{RN67}, secondary structure information \cite{RN68}, and the influence of metal ions \cite{RN69}, our approach aims to provide a holistic understanding of protein structures. This comprehensive approach underscores the interplay between chemical attributes and geometric features, contributing to a feature analysis of these essential biological features in macro-molecules.

\section{Methods}
\label{sec:method}
% We begin by exploring data augmentation, a pivotal step in ensuring the quality of our input data. Subsequently, we provide an in-depth look at our input format, defining how we represent protein graphs for our model's consumption. Finally, we introduce our innovative Chemical Residual Learning Framework, a critical constituent that integrates previously unexplored chemical features into the model to boost its capabilities.

This section will describe our protein data augmentation methods, NaNa and MiGu data augmentation, based on prior knowledge of biophysical and structural biology and the co-embedding residual learning framework. In \cref{subsec:overview}, we firstly give a high-level procedure of existing protein representation learning methods and our methods. In \cref{subsec:data_aug}, we will introduce our data augmentation algorithm for dynamic protein structural information as the first contribution. In \cref{subsubsec:node}, we will give further details of the side-chain biophysics and molecular interaction features for the node attribute augmentation. In \cref{subsubsec:edge_attributes}, we introduce chemical bonds as protein structural information as edge attributes. Furthermore, in \cref{subsubsec:migu_data_aug}, we combine the molecular interaction and side-chain biophysical features described in the previous section and propose two novel data augmentation methods, called NaNa and MiGu. In \cref{subsec:residual_learning_framework}, we introduce our second contribution, the residual learning framework, which can inject the augmented features into deeper layers to improve the protein classification tasks.

\subsection{Procedure Overview}
\label{subsec:overview}
% Our data augmentation pipeline is pivotal in enhancing the quality of input data for functional prediction tasks. It includes several key components:
In the beginning, we would abstract the procedure of existing works on protein classification tasks, consisting of four steps: (1) converting the PDB file to a graph, (2) enhancing the information with data augmentation based on graph datasets, and (3) combining the graph data, and (4) augmenting features to GNNs. Although existing works \cite{wang2022learning, wang2022comenet, schutt2018schnet} did not emphasize utilizing data augmentation method to enhance the dynamic structures, some of them \cite{wang2022learning, schutt2018schnet} incorporate naive data augmentation for dynamic structure augmentation and better performance. For instance, ProNet \cite{wang2022learning} leverages torsional angles perturbations as dynamic structure information, and ScheNet \cite{schutt2018schnet} introduces conformations variations to the learning framework. Furthermore, these methods did not amend lost segments in proteins. Therefore, we utilized moleculekit to fix the missing side-chain segments in proteins and incorporate chemical features from PropKa \cite{RN71} and AMBER \cite{case2005amber} to make the structural information accurate for the following processes.

\subsection{Data Augmentation}
\label{subsec:data_aug}

\subsubsection{Novel Node Attributes}
\label{subsubsec:node}
% \label{subsubsec:node_attr}
% \textbf{Node Coordinates} 
% \label{subsubsubsec:node_coord}
% We represent an amino acid with essential protein structural information from atoms, such as C$\alpha$, N, and O atoms extracted from PDB files, incorporating 9 three-dimensional Cartesian coordinates as features. 

\textbf{Molecular Biophysics Features}
\label{subsubsubsec:dynamic_feature}
% \SYNOTE{Re-Write}
% We utilized AMBER model and PropKa preprocessing to generate critical side-chain biophysical properties as chemical features \cite{RN71}. Additionally, AMBER model could fine-tune side-chain positions to obtain a more accurate structure according to empirical rule. This accurate simulation could provide a more realistic protonated environment. 
% For instance, the PropKa would yield the pKa value, which measures the balance of H+ and OH- in proteins, influencing protein structure and interactions due to charge-based forces. 
% Additionally, the charge-based forces would be simulated through the AMBER model, thus influencing the protein structures. Furthermore, PropKa and AMBER generate more features corresponding to dynamic structures of proteins, such as solvent accessibility, functional group types, and electrostatic states of each protein amino acid.

To achieve a more accurate semantic data augmentation of protein structures, we adopted the AMBER \cite{{case2005amber}} model and PropKa preprocessing \cite{RN71} to generate essential side-chain biophysical properties as chemical features, referencing \cite{RN71}. In this category, we extract side-chain chemical features, including pKa value, Functional groups, solvent accessibility, and electrostatic states. Specifically, the pKa value and electrostatic states are generated by PropKa \cite{RN71} and represent the ratio of H+ and OH- for better force field simulation. Moreover, with a better force field model of AMBER \cite{case2005amber}, we can build a more accurate protein structure and properties prediction because of more accurate attractions between molecules. Furthermore, functional groups and solvent accessibility in proteins are key determinants to determine the interaction in a protein \cite{solvant, function}. By augmenting that side-chain information, we could generate accurate semantic information to avoid unrealistic augmented protein structures.

\textbf{Secondary Structure Properties (SSP)} 
\label{subsubsubsec:dssp}
% (1) Why we need this feature?
% (2) Does it exist in the previous work? What's the difference
% (3) How can we incorporate it
% (4) What does the incorporated features can contribute?
% (5) Detail, sub-features
% To identify the comprehensive geometric features of protein backbone structures, we incorporate the protein secondary structure into our data augmentation, which is absent in the previous work. To achieve this goal, we leverage the geometric-based Defined Secondary Structure Properties (DSSP) algorithm to generate more accurate branch information for protein sequences based on Kabsch and Sander Theory \cite{kabsch1983dictionary}. Furthermore, we leverage the DSSP algorithm to explore three biochemistry features, including Secondary structure types (e.g., alpha-helix, beta-sheet), Significant geometric properties (e.g. phi and psi angles for each residue in the protein structures), and Solvent accessibility of each residue.

To synthesize more realistic and semantic protein backbone structures, we have incorporated the consideration of protein secondary structure into our data augmentation process, a component absent in previous studies \cite{wang2022learning, wang2022comenet, schutt2018schnet}. To achieve this, we utilize the geometry-based Defined Secondary Structure Properties (DSSP) algorithm, based on the Kabsch and Sander Theory \cite{kabsch1983dictionary}, to generate more accurate branching information for protein sequences. Additionally, we employ the DSSP algorithm to generate three biochemical features: secondary structure types (e.g., $\alpha$-helix, $\beta$-sheet), significant geometric properties (such as $\phi$ and $\psi$ angles for each residue in the protein structures), and solvent accessibility of each residue. Through the augmentation of these features, we could expand our understanding of protein structures. Secondary structure is a category system for protein sub-structure, leading to a more specific description of protein structures \cite{SS}. Furthermore, the protein $\phi$ and $\psi$ angles are the angles between molecules, which are parts of the protein structures. With such detailed structure information, we can augment more authentic protein context \cite{wang2022learning}.

\textbf{Node Type Features}
\label{subsubsubsec:node_type_feature}
Each node represents a C$\alpha$ atom of amino acid and metal ion. Metal ions and amino acids are the essential components of a protein, representing the important structural information and protein chemical components. We chose twenty-five widely existing amino acids in protein. Additionally, we also explore the contribution of metal ions and anions to the protein structure reconstruction, including Na$^+$, Mg$^{+2}$, Fe$^{+3}$, Cl$^-$, and SO$^{-4}$, based on the paper \cite{yamashita1990metal}. By incorporating the detailed node-type features, we can generate a more authentic protein structure. 

\subsubsection{Novel Edge Attributes}
\label{subsubsec:edge_attributes}
\textbf{Baker-Hubbard Theory} 
\label{subsubsubsec:baker}
Chemical bonds are important for performing biochemistry activities and stabilizing protein structure. Though the existing method incorporates geometric-based hydrogen bonds \cite{RN75}, their methods were not supported by a validated simulation model for extracting accurate bonding information. Therefore, to address inadequately represented the accurate semantic meaning of proteins, we utilized the Baker-Hubbard Theory \cite{hbond}, introducing an advanced bonding identification algorithm that surpasses previous methods in both depth and breadth of analysis. In addition, this method can be smoothly integrated into existing graph neuron network learning pipelines for data augmentation, thereby enriching datasets with comprehensive biochemical features and enhancing the learning capabilities of the models. 

\textbf{Empirical Binary Function} 
\label{subsubsubsec:empirical_bin_func}
Besides hydrogen bonds, there are four additional chemical bonds, including disulfide bonds, peptide bonds, $\pi$-$\pi$ interactions, and contacts within 8\r{A}.
We employ the Empirical Binary Function for bond data augmentation \cite{DiSu, pi}. Those functions calculate bond lengths, angles, and other vital characteristics, thus providing an accurate semantic dataset for algorithms to learn from, enabling them to recognize patterns and relationships in molecular data that were previously unexplored. 

% Based on Baker-Hubbard Theory and other chemical bonds \cite{baker1984hydrogen}, we leverage the bonding identification algorithm to explore two biochemistry features of hydrogen bonds.

% \textbf{Empirical Binary Function of bonds}
% \label{subsubsubsec:binary}
% The binary function of bonds is an empirical determination of the chemical bonds, including disulfide bonds, peptide bonds, $\pi$-$\pi$ interaction, and contacts (within 8\r{A}). This function involves calculating the length, angle, or relevant characteristics of the bond.

\subsubsection{Our Method: MiGu \& NaNa Data Augmentation}
\label{subsubsec:migu_data_aug}
We propose two novel semantic data augmentation methods: Novel Augmentation of New Node Attributes (NaNa) and Molecular Interactions and Geometric Upgrading (MiGu). The NaNa augmentation only consumes the node attributes in \cref{subsubsec:node}. On the other hand, we incorporate both node and edge attributes into MiGu augmentation in \cref{subsubsec:node} and \cref{subsubsec:edge_attributes}. We will evaluate both methods and present the results in \cref{subsec:impact_node_edge_attrs}.

Furthermore, our semantic augmentation method is computationally efficient. The augmentation process can be done within 4 seconds for each sample with only 3\% computational resource of Intel i7-9700K CPU and 614 MB memory. With proper parallel techniques, our methods can be applied to massive-scale datasets with little computation overheads compared to model training.

\subsection{Co-Embedding Residual Learning Framework}
\label{subsec:residual_learning_framework}
% \begin{figure*}[ht]
% \begin{center}
% \centerline{\includegraphics[width=\linewidth]{Residual_net_GIN.pdf}}
% \caption{Co-embedding Residual Learning Framework, This figure illustrates the detailed }
% \label{fig:figure 2}
% \end{center}
% \vskip -0.2in
% \end{figure*}

In this section, we introduce our second major contribution to the protein learning framework. Intuitively, to transfer the node and edge embeddings into deeper layers for better prediction accuracy, we incorporate the residual connection for information passing. The input format for a protein attributes are denoted as $\mathcal{G = }\left\{\mathcal{V},\mathcal{E}\right\}$, consists of several components: $\mathcal{V=}\left\{v_i\right\}_{i=1,\ldots,n}$ and $\mathcal{\mathcal{E}=}\left\{e_i\right\}_{i=1,\ldots,n}$ represents collections of node and edge attributes, where each $v_i,v_j \in \mathbb{R}^{n_c}$ and $e_i \in \mathbb{R}^{n_b}$ denote the feature vector for edge $i$, which is the edge between $v_i$ and $v_j$. Firstly, we made edge embeddings by concatenating two embeddings from adjacent nodes $v_{i} \oplus v_{j}$. Furthermore, we perform Hadamard product on node and edge embeddings $v_{i} \circ e_{i}$ in \cref{subsec:data_aug}. $L: \mathbb{R}^{n_{c}} \to \mathbb{R}^d$ is a trainable feature extractor of node and edge embeddings. In addition, we denote the output of layer $i$ as $u_i \in \mathbb{R}^{d_{i}}$ with corresponding to the $j$-th output entry as $u_{i}^j$ and the activation function as $\sigma: \mathbb{R}^{d_i} \to \mathbb{R}^{d_i}$.

Therefore, we design a co-embedding residual network presented as \cref{eq:residual}. In our model, we denote each layer of the neural network with an index $k$. We denote $L$ as the feature extractor described in \cref{sec:method}. Moreover, to support a comprehensive feature analysis in \cref{sec:method}, we denote $\mathcal{F}$ as any combinations of features described in $\mathcal{G}$. We also denote $\epsilon^{k-1}$ as a hyperparameter for balancing the information of graph representation and co-embedding attributes at the ${k-1}^{th}$ layer. Such architecture demonstrates faster training convergence and better prediction accuracy due to deeper node and edge information propagation shown in \cref{subsec:experiment_results}. 

% \PYB{Did we define $k$ anywhere?, Solved}

{\small
\begin{equation}
  \begin{split}
  u^k=\left(1+\epsilon^{k-1}\ \right)\cdot u^{k-1}+ \sigma\left(u^{k-1}+L (\mathcal{F})\right)
  \end{split}
  \label{eq:residual}
\end{equation}
}

\section{Experimental Design}

Our experimental design is multifaceted and aims to comprehensively evaluate the performance of various models when learning from data enriched with biophysical and chemical features. We have structured our experiments into three sections: implementation details in \cref{subsec:implementation_details} and datasets in \cref{subsec:datasets}.

\subsection{Implementation Details}
\label{subsec:implementation_details}
To evaluate our semantic data augmentation and co-embedding residual framework, we choose several GNNs as baseline models to compare the performance with and without our semantic augmentation and co-embedding residual framework. In our experiments, we choose GIN \cite{GIN}, GCN \cite{RN65}, and MPNN \cite{wang2022learning} as architecture baselines to evaluate the co-embedding learning framework due to their outstanding protein representation learning performance. Additionally, we also choose ProNet \cite{wang2022learning}, ComENet \cite{wang2022comenet}, and SchNet \cite{schutt2018schnet} 
% \PYB{wrong way of citation - should add citation right after the name, Solved} 
as the protein structure augmentation baselines to benchmark our semantic protein structure augmentation, NaNa, and MiGu.
% Our model training and optimization process incorporates several essential components seamlessly. We rely on the MPNN and GIN layer at the core of our GNN architecture. Furthermore, to enhance the model's capacity and adaptability, we integrate fully connected layers after each GNN layer, resulting in dimensions ranging from 64 to 128.
% Furthermore, we employ batch normalization (BatchNorm1d) within our model architecture to ensure stability during training and expedite convergence.

For model optimization, we turn to the Adam optimizer, a well-established choice for training networks.
During the training phase, we operate with a batch size of 256. Furthermore, for evaluation, we shifted to a batch size of 128. To fine-tune the learning process and facilitate optimal convergence, we employ a learning rate schedule that derives from 0.001 and gradually descends to 0.00001. We set the learning rate scheduler decay factor as 0.5 and the decay step as 60.
To address potential overfitting, we set universal setting dropout as 0.3, dimension size as 128, training batch size as 16, and validation batch size as 8 to test various frameworks in our experiments, allowing us to control the degree of regularization applied to the model precisely.

\subsection{Datasets}
\label{subsec:datasets}
Our research evaluates the efficacy of our methodology using two distinct datasets:

\subsubsection{SCOPe Classification Dataset}
SCOPe Classification Dataset \cite{SCOP} encompasses 12,312 training samples, 1,123 validation samples, and 710 testing samples. It primarily focuses on features related to protein structures in three-dimensional space and encompasses 1,123 different classes. Notably, this dataset is designed for fold, superfamily, and family classification tasks.

\subsubsection{EC Dataset}
Designed for predicting enzymatic functions, the EC dataset \cite{EC} comprises 29,210 training samples, 2,562 validation samples, and 5,645 testing samples. Both datasets exhibit balanced category distributions. The EC Dataset contains a total of 384 different classes.

% \subsubsection{DD Dataset}
% Designed for predicting enzymatic functions, the EC dataset comprises 1920 samples. The DD Dataset contains a total of 2  classes, divided into enzymatic and non-enzymatic properties. Because this dataset is lower in quantity, the performance is measured using k = 10-fold cross-validation.

\section{Experiment Results}
\label{subsec:experiment_results}
% In the initial phase of our experiments, we introduced a model that combines the Chemical Enhancement Module, Bonding Enhancement Module, and the Geometric Embedding module \cite{RN62}. These modules serve as extensions of our baseline models, encompassing a GCN-based or GIN-based baseline model. The models allow for the flexibility to include or exclude chemical features and the choice of incorporating or omitting chemical bonding information. Additionally, it provides the option to integrate a Bonding Enhancement Module to refine the analysis.
In the initial phase of our experiments, we introduced a model that combines the co-embedding residual learning framework. This learning framework serves as a component to enhance the learning of our baseline models, encompassing the GIN-based, GCN-based, and MPNN-based baseline models. In the second experiment, we incorporate node and edge attributes proposed in \cref{subsec:data_aug}, which are node attributes and edge attributes. Thirdly, to test the effectiveness of node attributes, we conduct leave-one-out feature analysis on the baseline model such as \cite{wang2022learning, wang2022comenet, schutt2018schnet}.

\vspace{-7mm}
\begin{figure}[htbp]
    % \vspace{-3mm}
    \begin{center}
    \centerline{\includegraphics[width=\linewidth]{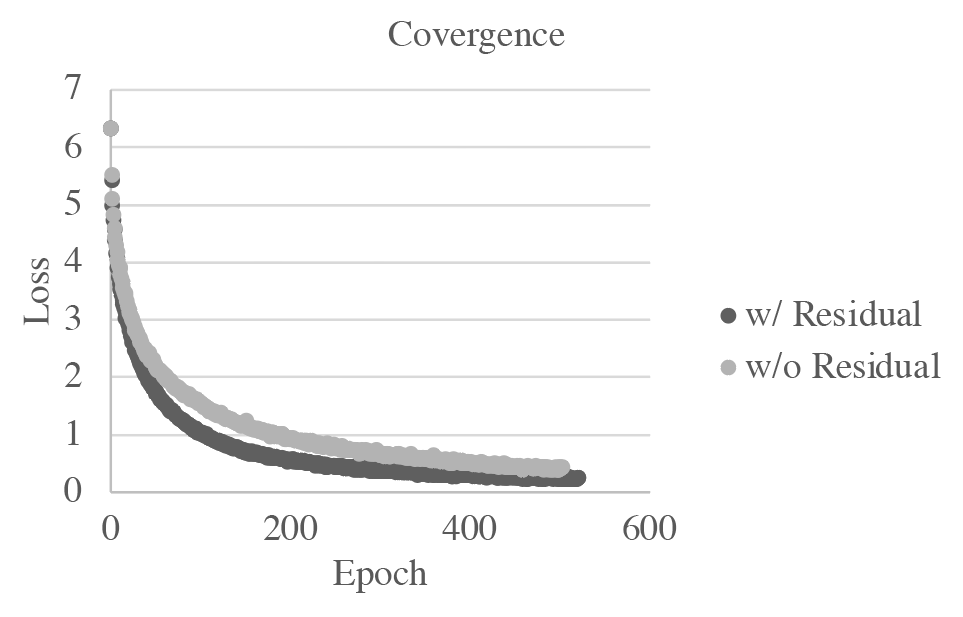}}
    \vspace{-5mm}
    \caption{This figure illustrates the difference in convergence speed between with and without residual learning framework on the EC dataset with the GIN model and NaNa semantic protein structure augmentation. The X-axis is the number of training epochs, and the Y-axis is the training loss. We can see that the convergence time of the model with residual learning framework can surpass the Vallina model without residual framework by 1.76 times.} 
    \label{fig:figure3}
    \end{center}
    \vskip -0.2in
    % \vspace{-15mm}
\end{figure}
\vspace{-5mm}
\subsection{The Effectiveness of Residual Learning Framework}
% {First of all, we would like to demonstrate 2 advantages of our co-embedding learning framework. Therefore, we incorporate the amino acid types as naive node attributes into the co-embedding residual learning framework and GNNs without residual frameworks. The first advantage is faster convergence speed. .... Additionally, the second advantage is better prediction accuracy....}
First of all, we would like to demonstrate two advantages of our co-embedding residual learning framework as \cref{tbl:res1} and \cref{fig:figure3}. Hence, we incorporate the amino acid types as naive node attributes, which are trivial information without affecting the performance, into the residual learning framework and compare them with the same models under naive model frameworks. Here, we choose GCN, GIN, and MPNN as baselines and evaluate them on the protein classification datasets, including EC, Fold, Superfamily, and Family datasets. The first advantage is faster training speed, in the \cref{fig:figure3}, we could see the convergence  become 1.76 
 times faster with the residual connection. Additionally, the second advantage of the residual framework is the better prediction accuracy. As shown in \cref{tbl:res1}, the residual framework could enhance the learning performance by at most 7.89\% accuracy like GCN with the Family dataset.

\subsection{Impact of Node and Edge Attributes}
\label{subsec:impact_node_edge_attrs}
In our second experiment, we explored the contributions of node and edge attributes to the performance enhancement independently. Therefore, in \cref{node}, we evaluate the accuracy of node and edge attributes independently on EC and SCOPs datasets with non-residual and residual architectures. Based on \cref{node}, we can see that our two semantic augmentations, NaNa and MiGu, significantly improve the accuracy of enzyme commission datasets by at most 16\% with residual learning framework. Additionally, in \cref{node}, the node attributes enhance more accuracy than edge attributes with residual frameworks. For example, the NaNa method (with Node attributes) outperforms the MiGu (with Node and Edge attributes) method by 4.12\% with the GIN and SuperFamily dataset. These results imply that the combination of co-embedding residual frameworks and node attributes can outperform the bonding information, which is a recognized crucial information in the biochemistry community. 

% In the second experiment, we aim to evaluate the effectiveness of node and edge attributes on prediction accuracy. Therefore, we evaluate the accuracy on the node and edge attribute independently in \cref{tbl:node_edge_impact}.
% \YSNOTE{This experiment aimed to elucidate the role of chemical bonding in enhancing the model's ability based on model performance.}
 % The GIN model with both node and edge attributes achieves the most significant improvement among the models in EC, Fold and Superfamily datasets, which are 80.90\%, 33.09\% and 49.05\% respectively. These findings underscore the positive contribution of node attributes to the protein property prediction. Nevertheless, the edge attributes contribute to a lower performance of co-embedding residual GIN model among EC, Fold and Superfamily datasets, signifying the deviation of traditional binary discern on the various bond types.
 % we found that incorporating \bold{only} node attributes could outperform the combination of node and edge attributes among EC, Fold and Superfamily datasets. Secondly, traditional binary discern on various bond types might be very inaccurate based on the first finding.  
% N of N 
%  N , which V O 
% A out perform B, AB comparable

\begin{table}[htbp]
    \vspace{-5mm}
    \setlength\tabcolsep{1pt}
    \caption{The Effectiveness of Residual Learning Framework. We demonstrate the improved performance of our residual learning framework, denoted as RES, across various Graph Neural Network models, such as GCN, MPNN, and GIN, on the EC and SCOPe datasets. The datasets are categorized by Fold, Superfamily, and Family levels of protein structure classification. We highlight the better result in each comparison in \textbf{bold}. In most test cases, the integration of residual connections leads to the best performance with at most 14\% improvement.
    } 
    % \YSNOTE{May require some summary for results? Solved}} \PYB{Yes please add, Solved} \PYB{this table should be divided into three parts by adding separation lines for each model; then for each part, you highlight Res as the best model}
    %GIN (RES) represents the GIN model combined with the residual connection layers. 
    % \begin{figure*}[ht]
    %     \begin{center}
    %     \centerline{\includegraphics[width=\linewidth]{}}
    %     \caption{Co-embedding Residual Learning Framework}
    %     \label{loss_conv}
    %     \end{center}
    % \vskip -0.2in
    % \end{figure*}
    
    \begin{center}
        \begin{small}
        \begin{sc}
        
            \begin{tabular}{lccccr}
            \toprule
            \label{tbl:res1}
            EXP1 & ~ & ~ & Dataset & ~ & ~ \\
            \toprule
            Model \quad \quad & EC \quad \quad & Fold \quad \quad & Super \quad \quad & Family \quad \\
            \midrule
             % GCN \quad\quad& 47.07 \quad\quad& 18.25 \quad\quad& 20.69 \quad\quad& 65.06\\
             GCN \quad\quad& 13.51 \quad\quad& 6.87 \quad\quad& 6.98 \quad\quad& 18.52\\
             % GCN(Res) & ~ & ~ & ~ & ~\\

             GCN(Res) \quad\quad& \textbf{18.09} \quad\quad& \textbf{9.27} \quad\quad& \textbf{11.27} \quad\quad& \textbf{26.41}\\       
            \midrule
             MPNN \quad\quad& 19.52 \quad\quad& 5.25
             \quad\quad& 10.86 \quad\quad& 24.45\\
             MPNN(Res) \quad\quad& \textbf{23.42} \quad\quad& \textbf{7.48} \quad\quad& \textbf{14.01} \quad\quad& \textbf{31.51}\\
            \midrule
             GIN \quad\quad& \textbf{64.79} \quad\quad& \textbf{22.36} \quad\quad& 31.47 \quad\quad& 83.31\\

             GIN(Res) \quad\quad& 64.49 \quad\quad& 21.76 \quad\quad& \textbf{37.38} \quad\quad& \textbf{88.73}\\

            % GRAPHSAGE & ~ & ~ & ~ & ~\\
            % GRAPHSAGE(Res) & ~ & ~ & ~ & ~\\

            \bottomrule
            \end{tabular}
        % \label{tbl:res}
        \end{sc}
        \end{small}
    \end{center}
    \vskip -0.1in
    \vspace{-5mm}
\end{table}

\begin{table}[htbp]
    \vspace{-2mm}
    \caption{Performance Comparison. We compare our methods, NaNa and MiGu, with and without residual framework using EC, Fold, Superfamily, and Family datasets. The best results are \textbf{bolded}, and the second-best results are indicated with a \underline{slash}. RES label represents the various models combined with the residual connection layers. NaNa augmentation results outperform most test cases among three baseline models of GIN, GCN, and MPNN. Additionally, MiGu augmentation demonstrates the second-highest performance across various models and metrics. 
    % \YSNOTE{some summaries for results} \PYB{Need to rescale the table to fit into one column, Solved} 
    }
    
    \label{node}
    \vskip 0.3\in
    \vspace{-5mm}
    \begin{center}
        \begin{small}
            \begin{sc}
            \resizebox{\linewidth}{!}{
                \begin{tabular}{lccccr}
                \toprule
                EXP2 & ~ & ~ & Dataset & ~ & ~ \\
                \toprule
                AUG & Model & EC & Fold & Super &Family \\
                % \midrule
                % & GCN(Res) & 18.09 & 9.27 & 11.27 & 26.41\\

                % & MPNN(Res) & 23.42 & 7.48 & 14.01 & 31.51\\
                % & GIN(Res) & 64.49 & 21.76 & 37.38 & 88.73\\
                \midrule
                % Nana & GCN & 65.43 & 24.97 & 30.70 & 78.23\\

                Nana & GCN & 16.54 & 9.10 & 8.97 & 19.77\\
                w/o & GCN(Res) & 21.34 & 9.27 & 11.27 & 26.41\\
                %13 20 37
                Nana & GCN(Res) & \underline{26.78} & \underline{14.13} & \underline{20.61} & \underline{39.22}\\

                % N&GIN(res) & 73.01&26.99&40.55&91.28\\
                MiGu & GCN(Res) & \textbf{27.26} & \textbf{14.86} & \textbf{25.00} & \textbf{42.03}\\  

                \midrule
                Nana & MPNN & 18.80 & 10.23 & 17.03 & 30.89\\
                %ok
                w/o& MPNN(Res) & 23.42 & 7.48 & 14.01 & 31.51\\
                Nana & MPNN(Res) & \textbf{26.94} & \underline{11.01} & \underline{19.10} & \underline{33.20}\\

                % N&GIN(res) & 73.01&26.99&40.55&91.28\\
                %ok
                MiGu & MPNN(Res) & 

                \underline{24.96} & \textbf{12.92} & \textbf{22.38} & \textbf{38.44}\\
                \midrule
                Nana & GIN & 73.01&26.99&40.55&91.28\\
                w/o & GIN(Res) & 64.49 & 21.76 & 37.38 & 88.73\\
                Nana&\textbf{GIN(res)}    & \textbf{80.90} & \textbf{33.09} & \textbf{49.05} & \underline{92.31}\\

                MiGu& \textbf{GIN(res)} & \underline{77.34} & \underline{31.02} & \underline{44.93} & \textbf{92.86}\\

                \bottomrule
                \end{tabular}
                }
            \end{sc}
        \end{small}
    \end{center}
    \vskip -0.1in
    \vspace{-5mm}
\end{table}
% \usepackage{float}
% \vspace{-50mm}
\begin{table}[htbp]
    % \vspace{-15mm}
    \caption{Performance Comparison of Leave-One-Out Feature Analysis. We ablate various features, biophysical features, SSP, and node-type features using Fold, Superfamily, and Family datasets. The best results are \textbf{bolded}, and the second-best results are indicated with a \underline{slash}. We found that all kinds of features could enhance the performance on various baselines. Specifically, the DSSP enhances the performance the most.}
    \vspace{-5mm}
    \begin{center}
        \begin{small}
        % \begin{sc}
        \resizebox{\linewidth}{!}{
            \begin{tabular}{lccccr}
            \toprule
            \label{tbl:baseline_compa}

            EXP3 & ~ & ~ & Dataset & ~ & ~ \\
            \toprule
            ~ & Model & EC& Fold & Super & Family \\
            \midrule
            Original & SchNet & 53.12 & 18.11 & 21.85 & 76.42 \\
            ~ &ComENet & 70.39 & 27.02 &40.51 &92.14\\
            ~ &ProNet & 79.63&45.68&60.05&\textbf{97.41}\\
            % \\
            w/o BioPhys & SchNet &66.15&32.40&43.45&85.98 \\
            ~ &ComENet &80.03&38.41&54.47&94.56\\
            ~ &ProNet & \underline{82.81}&46.23&\textbf{63.58}&96.85\\
            % \\
            w/o SSP & SchNet &57.27&22.35&25.00&79.37 \\
            ~ &ComENet &72.41&28.07&39.30&90.79\\
            ~ &ProNet & 80.99&46.93&60.53&\underline{97.32}\\
            % \\
            % w/o DSSP & SchNet &~&~&~&~ \\
            % ~ &ComENet &~&~&~&~\\
            % ~ &ProNet & ~&~&~&~\\
            
            w/o Node Type & SchNet &63.49&31.70&40.10&84.65 \\
            ~ &ComENet &81.45&37.51&53.43&95.20\\
            ~ &ProNet & 82.35&\underline{48.32}&\underline{63.42}&97.24\\
        
            NANA & SchNet & 66.12&31.28&43.13&88.27\\
            ~  & ComENet & 82.01&40.64&57.11&95.83\\
             ~  & ProNet  & \textbf{83.95} &\textbf{48.46} &61.98&\underline{97.32}\\

            \bottomrule
            \end{tabular}
        }
        % \end{sc}
        \end{small}
    \end{center}
    % \label{tbl:baseline_compa}
    \vskip -0.1in
\end{table}

\subsection{Leave-One-Out Feature Analysis}
The primary objective of this experiment was to assess the effectiveness of sub-features of node attributes in various baseline models. 
We conduct Leave-One-Out feature analysis on various node features among three baselines. 

Surprisingly, when excluding the node type features, we found the performance sometimes remained the same as comprehensive feature incorporation for ProNet \cite{wang2022learning}. However, we found that the node-type features can usually bring prediction improvement among all kinds of models and datasets, including ComENet, and ScheNet. Shown in \cref{tbl:baseline_compa},  the node type features bring improvement for ComENet on Fold and Superfamily datasets from 37.51\% to 40.64\% and 53.43\% to 57.11\% respectively.

Furthermore, we found that the DSSP features could bring significant improvement in various baselines for different datasets. Specifically, the DSSP features could enhance the performance on ComENet from 37.51\% to 40.64\% and 53.43\% to 57.11\% on Fold and Superfamily datasets. 

In addition,  We found that dynamic features could slightly improve the ComENet baseline. For example, the dynamic features enhance the performance from 1\% to 3\% among various datasets.

\subsection{Influence of Node Features}
\label{subsec:infl_node_features}

In our third experiment, we conducted a comparative analysis of the accuracy contribution of node attributes. We performed this assessment on both the EC \cite{EC} and SCOPe datasets \cite{SCOP} to understand how these additional sources of information affect model performance. 
Among these results, our augmentation method could improve state-of-the-art model architectures on both the EC and SCOP datasets. The results are presented in \cref{tbl:baseline_compa}. 
% By comparing models trained with and without node attributes, we gained insights into the significance of chemical-supplemented data in improving model accuracy in the context of protein property prediction. \ref{tbl:baseline_compa} represents our experimental results. \YSNOTE{table number wrong}

We found that the node attributes could bring significant improvements on various baselines and datasets, surpassing the state-of-the-art model like ProNet \cite{wang2022learning}.
Shown in table \cref{tbl:baseline_compa}, we found that our method, NaNa data augmentation, could get at least 4\% improvement on EC datasets in all baselines. Furthermore, the node attributes could enhance the Fold and Superfamily datasets by at least 3\% and 1\%, respectively.  
We assessed the impact of integrating node attributes, including node types, DSSP, and dynamic features. In the EC dataset, the baseline model with node attributes information achieved an accuracy of 83.95\%, succeeding the performance of ProNet without node attributes, highlighting the effects of node attributes. Similarly, in the Fold dataset, the model with comprehensive node attributes achieved significant accuracy scores, reaching 48.46\% on Fold testing data, 61.98\% on superfamily testing data, and 97.32\% on family testing data, which comparatively surpassed the performance of ProNet without node attributes, which are 45.68\%, 60.05\%, and 97.41\%. This again emphasizes the importance of node types, DSSP, and dynamic features in enhancing model performance.

This result emphasizes that our comprehensive integration of residual networks led to exceptional performance on both EC and Fold datasets, surpassing the baseline models. 

\vskip -0.1in
\vspace{-1mm}
\section{Conclusion}
\label{sec:conclusion}
This paper proposes novel semantic protein structure data augmentation techniques based on biophysic prior knowledge and an effective residual architecture for protein representation learning, bringing significant improvement to protein classification tasks. In addition, our work provides comprehensive feature analysis and surpasses the state-of-the-art baseline on protein classification graph models with the augmentation of biophysical, SSP, amino acid, and ionic types features.

Furthermore, we develop a co-embedding residual network with biochemistry and dynamic geometric features, which could apply to the GIN model with a shorter convergence time for improved training and better test accuracy.

% In summary, we provide a pilot study on incorporating dynamic structure and biochemistry information into protein properties prediction. 

Our results shed new light on the importance of incorporating biophysical features to improve machine learning in protein classification tasks and a corresponding architecture that can effectively extract the augmented features.

\nocite{langley00}

\bibliography{mine}
\bibliographystyle{icml2023}

%%%%%%%%%%%%%%%%%%%%%%%%%%%%%%%%%%%%%%%%%%%%%%%%%%%%%%%%%%%%%%%%%%%%%%%%%%%%%%%
%%%%%%%%%%%%%%%%%%%%%%%%%%%%%%%%%%%%%%%%%%%%%%%%%%%%%%%%%%%%%%%%%%%%%%%%%%%%%%%
% APPENDIX
%%%%%%%%%%%%%%%%%%%%%%%%%%%%%%%%%%%%%%%%%%%%%%%%%%%%%%%%%%%%%%%%%%%%%%%%%%%%%%%
%%%%%%%%%%%%%%%%%%%%%%%%%%%%%%%%%%%%%%%%%%%%%%%%%%%%%%%%%%%%%%%%%%%%%%%%%%%%%%%
\newpage
\appendix
\onecolumn

%%%%%%%%%%%%%%%%%%%%%%%%%%%%%%%%%%%%%%%%%%%%%%%%%%%%%%%%%%%%%%%%%%%%%%%%%%%%%%%
%%%%%%%%%%%%%%%%%%%%%%%%%%%%%%%%%%%%%%%%%%%%%%%%%%%%%%%%%%%%%%%%%%%%%%%%%%%%%%%

\end{document}